\documentclass[twoside,twocolumn,9pt]{article}
\usepackage{extsizes}
\usepackage[super,sort&compress,comma]{natbib} 
\usepackage[version=3]{mhchem}
\usepackage[left=1.5cm, right=1.5cm, top=1.785cm, bottom=2.0cm]{geometry}
\usepackage{balance}
\usepackage{mathptmx}
\usepackage{sectsty}
\usepackage{graphicx} 
\usepackage{lastpage}
\usepackage[format=plain,justification=justified,singlelinecheck=false,font={stretch=1.125,small,sf},labelfont=bf,labelsep=space]{caption}
\usepackage{float}
\usepackage{fancyhdr}
\usepackage{fnpos}
\usepackage[english]{babel}
\addto{\captionsenglish}{%
  
}
\usepackage{array}
\usepackage{droidsans}
\usepackage{charter}
\usepackage[T1]{fontenc}
\usepackage[usenames,dvipsnames]{xcolor}
\usepackage{setspace}
\usepackage[compact]{titlesec}
\usepackage{hyperref}

\usepackage{epstopdf}

\definecolor{cream}{RGB}{222,217,201}

\begin{document}

\pagestyle{fancy}
\thispagestyle{plain}
\fancypagestyle{plain}{
\renewcommand{\headrulewidth}{0pt}
}

\makeFNbottom
\makeatletter
\renewcommand\LARGE{\@setfontsize\LARGE{15pt}{17}}
\renewcommand\Large{\@setfontsize\Large{12pt}{14}}
\renewcommand\large{\@setfontsize\large{10pt}{12}}
\renewcommand\footnotesize{\@setfontsize\footnotesize{7pt}{10}}
\makeatother

\renewcommand{\thefootnote}{\fnsymbol{footnote}}
\renewcommand\footnoterule{\vspace*{1pt}%
\color{cream}\hrule width 3.5in height 0.4pt \color{black}\vspace*{5pt}} 
\setcounter{secnumdepth}{5}

\makeatletter 
\renewcommand\@biblabel[1]{#1}            
\renewcommand\@makefntext[1]%
{\noindent\makebox[0pt][r]{\@thefnmark\,}#1}
\makeatother 
\renewcommand{\figurename}{\small{Fig.}~}
\sectionfont{\sffamily\Large}
\subsectionfont{\normalsize}
\subsubsectionfont{\bf}
\setstretch{1.125} 
\setlength{\skip\footins}{0.8cm}
\setlength{\footnotesep}{0.25cm}
\setlength{\jot}{10pt}
\titlespacing*{\section}{0pt}{4pt}{4pt}
\titlespacing*{\subsection}{0pt}{15pt}{1pt}

\fancyfoot{}
\fancyfoot[LO,RE]{\vspace{-7.1pt}\includegraphics[height=9pt]{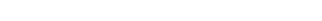}}
\fancyfoot[CO]{\vspace{-7.1pt}\hspace{13.2cm}\includegraphics{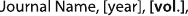}}
\fancyfoot[CE]{\vspace{-7.2pt}\hspace{-14.2cm}\includegraphics{head_foot/RF}}
\fancyfoot[RO]{\footnotesize{\sffamily{1--\pageref{LastPage} ~\textbar  \hspace{2pt}\thepage}}}
\fancyfoot[LE]{\footnotesize{\sffamily{\thepage~\textbar\hspace{3.45cm} 1--\pageref{LastPage}}}}
\fancyhead{}
\renewcommand{\headrulewidth}{0pt} 
\renewcommand{\footrulewidth}{0pt}
\setlength{\arrayrulewidth}{1pt}
\setlength{\columnsep}{6.5mm}
\setlength\bibsep{1pt}

\makeatletter 
\newlength{\figrulesep} 
\setlength{\figrulesep}{0.5\textfloatsep} 

\newcommand{\topfigrule}{\vspace*{-1pt}%
\noindent{\color{cream}\rule[-\figrulesep]{\columnwidth}{1.5pt}} }

\newcommand{\botfigrule}{\vspace*{-2pt}%
\noindent{\color{cream}\rule[\figrulesep]{\columnwidth}{1.5pt}} }

\newcommand{\dblfigrule}{\vspace*{-1pt}%
\noindent{\color{cream}\rule[-\figrulesep]{\textwidth}{1.5pt}} }

\makeatother

\twocolumn[
  \begin{@twocolumnfalse}
{\includegraphics[height=30pt]{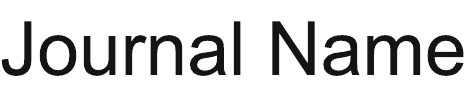}\hfill\raisebox{0pt}[0pt][0pt]{\includegraphics[height=55pt]{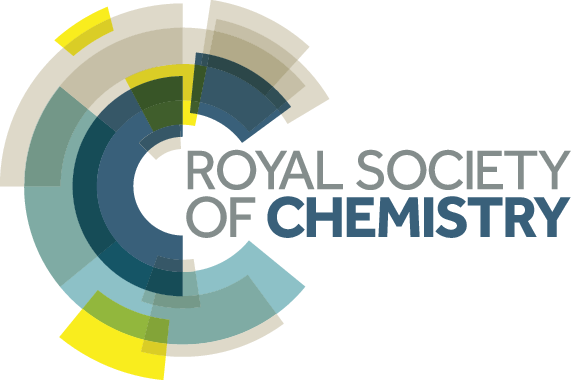}}\\[1ex]
\includegraphics[width=18.5cm]{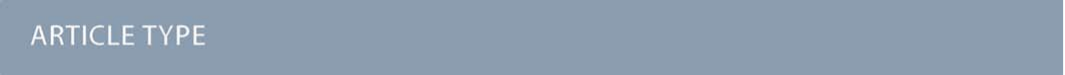}}\par
\vspace{1em}
\sffamily
\begin{tabular}{m{4.5cm} p{13.5cm} }

\includegraphics{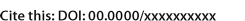} & \noindent\LARGE{\textbf{Low-energy pathways lead to self-healing defects in CsPbBr$_3$$^\dag$}} \\
\vspace{0.3cm} & \vspace{0.3cm} \\

 & \noindent\large{Kumar Miskin,$^{\ast}$\textit{$^{a}$} Yi Cao,\textit{$^{b}$} Madaline Marland,\textit{$^{b}$} Jay I. Rwaka,\textit{$^{b}$} Farhan Shaikh,\textit{$^{b}$} David T. Moore,\textit{$^{c}$} John A. Marohn\textit{$^{d}$} and Paulette Clancy \textit{$^{b}$}} \\

\includegraphics{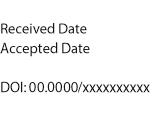} & \noindent\normalsize{Self-regulation of free charge carriers in perovskites via Schottky defect formation has been posited as the origin of the well-known defect tolerance of metal halide perovskite materials that are promising candidates for photovoltaic applications, like solar cells.  
Understanding the mechanisms of self-regulation, here for a representative of more commercially viable all-inorganic perovskites, promises to lead to the fabrication of better-performing solar cell materials with higher efficiencies.
We investigated different mechanisms and pathways of the diffusion and recombination of interstitials and vacancies (Schottky pairs) in \ce{CsPbBr3}. 
We use Nudged Elastic Band calculations and \textit{ab initio}-derived pseudopotentials within Quantum ESPRESSO to determine energies of formation, migration, and  activation for these defects. 
Our calculations uncover defect pathways capable of producing an activation energy at or below the value of 0.53~eV observed for the slow, temperature-dependent recovery of light-induced conductivity in \ce{CsPbBr3}. 
Our work reveals the existence of a low-energy diffusion pathway involving a concerted ``domino effect'' interstitial mechanism, with the net result that interstitials can diffuse more readily over long distances than expected. 
Importantly, this observation suggests that defect self-healing can be promoted if the ``domino effect'' strategy can be engaged. } \\

\end{tabular}

 \end{@twocolumnfalse} \vspace{0.6cm}

  ]

\renewcommand*\rmdefault{bch}\normalfont\upshape
\rmfamily
\section*{}
\vspace{-1cm}


\footnotetext{\textit{$^{a}$~Department of Materials Science and Engineering, Johns Hopkins University, USA.  E-mail: kmiskin1@jh.edu}}
\footnotetext{\textit{$^{b}$~Department of Chemical and Biomolecular Engineering, Johns Hopkins University, USA }}
\footnotetext{\textit{$^{c}$~National Renewable Energy Lab, USA}}
\footnotetext{\textit{$^{d}$~Department of Chemistry and Chemical Biology, Cornell University, USA}}

\footnotetext{\dag~Electronic Supplementary Information (ESI) available: See DOI: 00.0000/00000000.}


\section{Introduction}

Metal halide perovskites (MHPs) are one of the most exciting materials classes at the forefront of photovoltaic materials development.\ 
They have demonstrated performance comparable to market-leading solar cell materials, as evidenced by their dramatic rise in power conversion efficiency (PCE) since emerging about a decade ago.\cite{Manser2016IntriguingPerovskites}\
MHPs, like all perovskites, adopt the general formula \ce{ABX3}, where A is a monovalent cation (\ce{MA^+}, \ce{FA^+}, \ce{Cs^+}), B is a divalent metal cation (\ce{Sn^2+}, \ce{Pb^2+}), and X is a halide (\ce{Br^-}, \ce{I^-}, \ce{Cl^-}).\ Early work on MHPs centered on methylammonium (MA) lead iodide (or \ce{MAPbI3}). More recently, the field is migrating towards all-inorganic perovskites, like CsPbX$_3$, with band gaps in the range 1.7-2.3 depending on the halide. These perovskites have proven to be somewhat more durable in conditions (UV, heat, and humidity) that can degrade \ce{MAPbI3} and other hybrid organic-inorganic perovskites over time.  Due to its importance, we focus our defect studies on \ce{CsPbBr3}.\

Defect migration and recombination strongly affect device performance and long-term durability and are generally unwanted in photovoltaic devices. \cite{Yuan2016IonStability}\
In that regard, metal halide perovskites (MHP) are known to be far more defect-tolerant than their silicon solar cell counterparts.\cite{Yin2014UnusualAbsorber,Walsh2015SelfRegulationPerovskites,Kang2017HighCsPbBr3}  
Surprisingly, the underlying atomic-level mechanisms and prevalent pathways for defect creation and remediation remain largely unknown. \cite{Brandt2017SearchingScreening}\ 
The most widely accepted theory posits that the most probable perovskite defects form ``shallow'' traps near the conduction or valence band and therefore still allow long lifetimes for charge carriers. \cite{Shi2014ShallowMaterials, Kang2017HighCsPbBr3}\
Conversely, experimental work argues that deeper traps must be present \cite{Heo2017DeepSpectroscopy} and that ionic defects, which may be created by light \cite{Kim2018LargePhotodecomposition,Senocrate2019Solid-StatePerovskites} and voltage,\cite{Senocrate2019Solid-StatePerovskites} are a source of device hysteresis and instability.\cite{Kim2021PhotoEffectPhotoDemixing,Moia2021IonCells}

Experimental measurements of defect activation energies are rare, especially for the all-inorganic perovskite studied in this paper. Here, we obtain activation energies, E$_a$,  computationally: The activation energy is determined from the sum of the energy to \textit{form} the defect, E$_f$,  plus the energy to \textit{move} the defect in the lattice, E$_m$.  Both E$_f$ and E$_m$ can be determined computationally, here using density functional theory calculations, and hence activation energies can be estimated. 

In terms of extant defect studies, especially computational studies, a larger volume of prior work exists for hybrid organic-inorganic systems, with methylammonium (\ce{MA+}) and formamidinium (\ce{FA+}) A-site cations in the lattice, compared to fewer studies for \textit{inorganic} MHP systems. \cite{Azpiroz2015DefectOperation, Minns2017StructureIodide, Xue2021First-principlesFunctionals}\
Meggiolaro and De Angelis provided a fairly comprehensive review of \textit{ab initio}-generated formation energies for \ce{MAPbI3} and offered ``best practices'' for such calculations. \cite{Meggiolaro2018First-PrinciplesIssues} 
Oranskaia \textit{et al.} found lower hydrogen-bonding strength in \ce{MAPbBr3} than \ce{FAPbBr3} and, consequentially, lower halide migration and defect formation energies (DFE). \cite{Oranskaia2018HalogenMatters}\
This finding suggests that careful A-site cation selection may reduce perovskite ion migration. 

While the available literature reports defect energies, mechanistic understanding is currently limited, especially for all-inorganic MHP materials.  
Evarestov \textit{et al.} found that perfect \ce{CsPbX3} structures follow the stability trend \ce{CsPbBr3} > \ce{CsPbI3} > \ce{CsPbCl3}, corroborating \ce{CsPbBr3} defect formation energies from Kang and Wang. \cite{Evarestov2020First-principlesCrystals,Kang2017HighCsPbBr3}\ 
In \ce{CsPbI3} systems, Xue \textit{et al.} found that an anti-site defect of Pb on a Cs site forms with comparable ease  to other defects. \cite{Xue2023CompoundCsPbI3} 
Balestra reported significantly lower activation energy for vacancy migration for mixed halide \ce{CsPb(Br_{x}I_{1-x})3} ($x = 1/3$, $2/3$) than for homogeneous halide systems in which all three X atoms are the same halide. \cite{Balestra2020EfficientPerovskites}\
Ten Brinck \textit{et al.} calculated defect formation energy values for interstitial, vacancy, and anti-site defects in non-periodic 3~$nm$ \ce{CsPbBr3} nanoparticles and concluded that the
 most stable defect position is on the nanoparticle surface. \cite{TenBrinck2019DefectsBulk}

Our calculations are motivated by Tirmzi \emph{et al.}'s measurements of the slow recovery of light-induced conductivity in \ce{CsPbBr3}.\cite{Tirmzi2017CoupledPerovskite}
In these measurements, a charged atomic-force microscope cantilever was brought near the surface of a \ce{CsPbBr3} film and experienced a non-contact friction depending on the sample's total conductivity, electronic plus ionic.\cite{Tirmzi2019Substrate-DependentAlloy}
After irradiating the sample with above-bandgap light, Tirmzi \emph{et al.}\ observed a prompt increase in friction, as expected from the presence of photo-generated electrons and holes.
Electrons and holes recombine on the microsecond timescale in perovskites.
After the light was turned off, however, sample friction took $10$ s to recover at room temperature.
Tirmzi \emph{et al.}\ concluded that the sample's conductivity was dominated by ionic conductivity that that light was \emph{creating} mobile ionic defects:
\begin{equation}
  \mathrm{nil} 
    \ce{->[$h \nu$]}
    \mathrm{Br}_\mathrm{Br}^\mathrm{x} 
      + \mathrm{h}^{\bullet} 
      + e^{\prime}
    \ce{->}
    \mathrm{V}_\mathrm{Br}^{\bullet} 
      + \mathrm{Br}_\mathrm{i}^\mathrm{x}
      + e^{\prime}, 
    \label{eq;1}
\end{equation}
in Kr\"{o}ger-Vink notation.
The friction recovery time in the dark was temperature dependent, with an activation energy of $E_{\mathrm{a}} = 0.53 \pm 0.03 \: \mathrm{eV}$ (95\% confidence interval).
The associated defect-recovery reaction in the dark is
\begin{equation}
    \mathrm{V}_\mathrm{Br}^{\bullet} 
      + \mathrm{Br}_\mathrm{i}^\mathrm{x}
      + e^{\prime}
     \ce{->} \mathrm{nil}.
    \label{eq;1b}
\end{equation}
For this reaction we would expect the measured $E_{\mathrm{a}}$ to be dominated by the \emph{migration} energy governing the diffusion of the neutral interstitial $\mathrm{Br}_\mathrm{i}^\mathrm{x}$.

In this paper, we have modeled a number of potential recombination pathways for interstitial, vacancy, and anti-site defects using Nudged Elastic Band -- Density Functional Theory (NEB-DFT).\
Unlike previous studies in the literature, we have calculated both formation and migration energies for a range of defect pathways, allowing us the ability to calculate activation energies, an experimentally accessible property. 
Coupling Nudged Elastic Band calculations with \textit{ab initio} pseudopotentials provides accurate estimates for the migration energies.  
The conductivity-recovery rate observed by Tirmzi and coworkers \cite{Tirmzi2017CoupledPerovskite} was not noticeably different near grain boundaries; they were observing a bulk effect.
This observation justifies our ignoring ground-boundary effects in the following computations.

We are looking to find a match between computational and experimental activation and migration energies and, as a result, uncover the dominant defect(s) and defect pathways that are present.
With this information in hand, we hope to better inform device synthesis and offer a different bulk-defect passivation strategy to improve overall MHP performance.\

\section{Results and Discussion}

\begin{figure*}
    \centering
    \includegraphics[width=\textwidth]{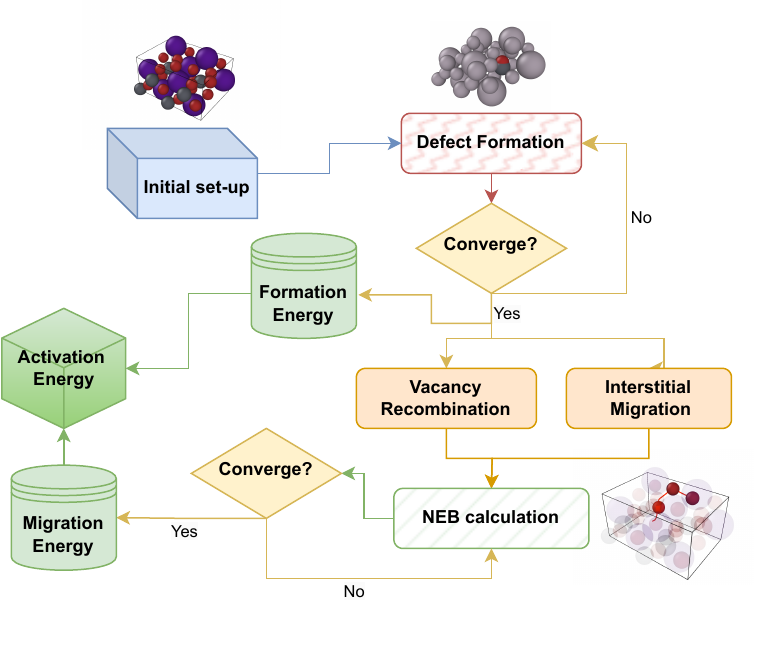}
    \caption{Flowchart of defect studies in \ce{CsPbBr3}.
    This begins with an initial lattice set-up (top left; blue box).
    Then a defect is created, whether a vacancy or an interstitial (top right; pink box). 
    Converged calculations yield a formation energy. Additional calculations using NEB produce the migration energy (bottom right; green box), for a set of considered pathways (middle; orange boxes). Converged NEB calculations produce E$_m$. Then the activation energy can be calculated (dark green cube).}
    \label{fig:flowchart}
\end{figure*}

To validate our NEB calculations, we first predicted an X-ray diffraction (XRD) spectrum generated from our simulated pristine supercell and VESTA software\cite{Momma2011VESTAData} and compared it against that found experimentally for a thin film MHP by our collaborators.
As seen in the comparison, found in the Supplementary Information, the simulated peaks line up essentially perfectly, with the density functional theory-generated observations matching the experimental results.
The relaxed cell parameters for the pristine \ce{CsPbBr_3} supercell ($3 \times 3 \times 2$) (with \textit{Pnma} spacegroup) were found to be 8.267, 8.177 and  11.768 \r{A}. 
These correspond very favorably to experimentally reported values:\cite{Lopez2020CrystalSynthesis} 8.23, 8.18 and 11.73 \r{A}.

\subsection{Defect Formation Energies}

First, we examined defect formation energy values in \ce{CsPbBr3}, focusing on the consideration of various kinds of interstitial sites, as mentioned in the Methods section. 
The \ce{Br_i} atom was placed in a chosen interstitial position and allowed to relax in the crystal.
We explored different Br interstitials based on their local electronic environment as explained in more detail later in Section~\ref{defect_migration}.

For the \ce{Br_i-Br_i} split-interstitial system, a Br interstitial shares a lattice site with another Br atom. For the \ce{Br_i-Pb_i} split-interstitial system, a Br interstitial shares a lattice site with a Pb atom. 
The defect formation energy (DFE) for this configuration varies depending on the axis along which the Pb-Br pair is aligned. 
As depicted in Figure S2 of the Supplementary Information, the DFE is found to be \(0.83\)~eV, on average, when the pair is aligned along the $z$-axis. 
In contrast, the DFE value increases to \(1.14\)~eV when the split interstitial is aligned along the $y$-axis. 
This difference of approximately 0.4~eV between the two orientations is significant and consistent, indicating a pronounced anisotropic phenomenon in defect formation.

In the \ce{Pb_{Br}} and \ce{Br_{Pb}} anti-site systems, the defect formation energy varied with the location of the introduced anti-site. 
It is observed that, as the anti-site is placed closer to the center of the supercell system, the DFE comes closer to ten Brinck's experimental values. \cite{TenBrinck2019DefectsBulk}
This can be explained by the effect of periodic boundary conditions on the simulation, wherein it is observed that defects near the edges of the supercell tend to result in distorted energy values. 
All the values are reported in the Supplementary Information. 
Our calculations resulted in large formation energies of 6.08~eV and 1.30~eV for the \ce{Pb_{Br}} and \ce{Br_{Pb}} anti-site defects, respectively. These values are similar to those reported by Kang \textit{et al.} \cite{Kang2017HighCsPbBr3} 
\ce{Pb_{Br}} shows a high DFE due to the introduction of a larger Pb atom in place of a smaller Br, causing a considerable shift in neighboring atomic positions.

Table \ref{tab:table1} reports DFE values for all the defects that were studied in this work. 

\subsection{Defect Migration Energies} \label{defect_migration}
\renewcommand{\thesubsubsection}{\arabic{section}.\arabic{subsection}.\arabic{subsubsection}}

\begin{figure*}
    \centering
    \includegraphics[width = \textwidth]{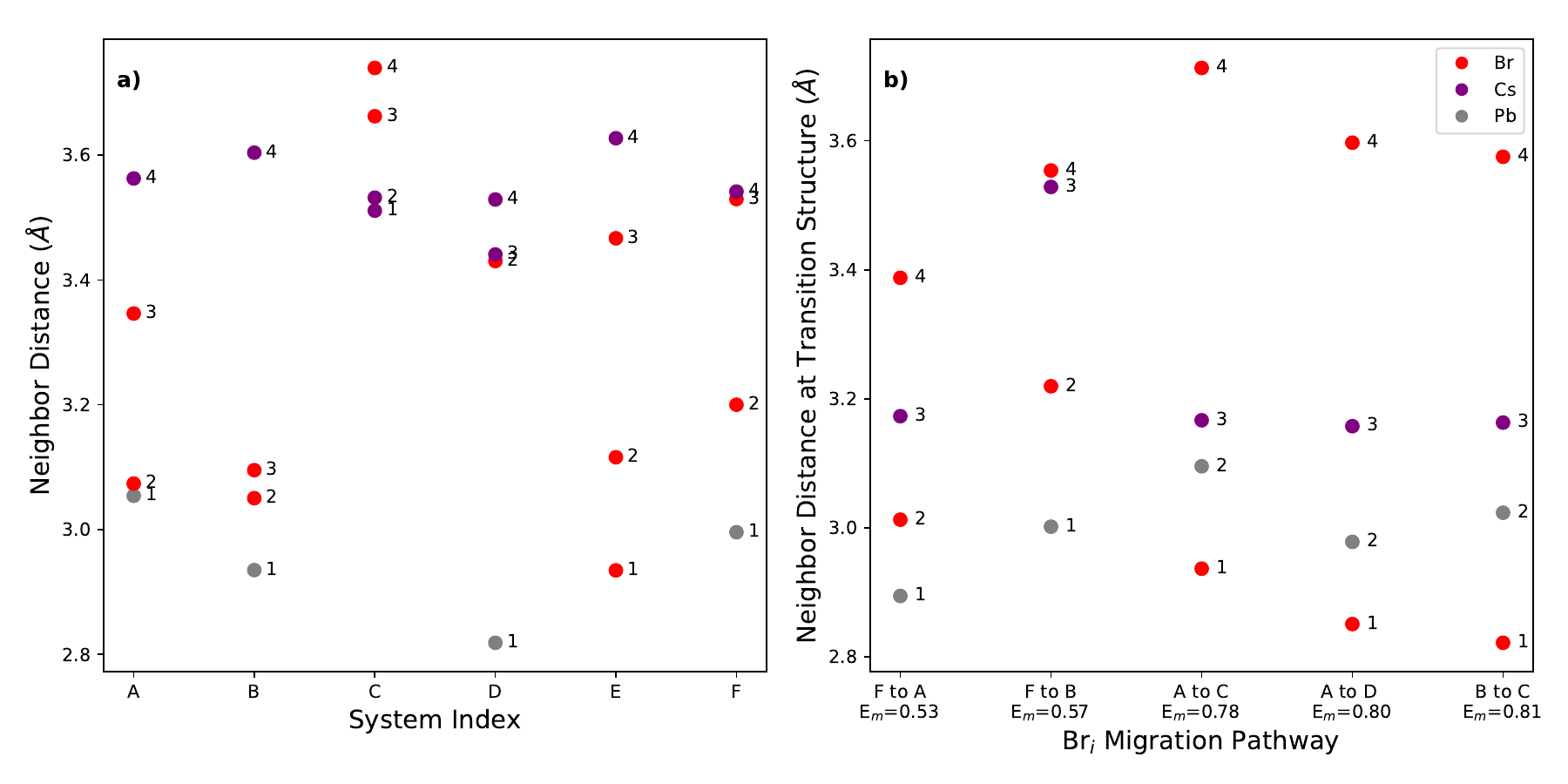}
    \caption{Locations of first, second, third, and fourth neighbor interatomic distances (Å) from a reference neutral interstitial Br$_i$ in (a) relaxed $2 \times 1 \times 1$ \ce{CsPbBr3} structures and (b) neighbor distances found at the NEB transition state, sorted from lowest E$_m$ to highest on the $x$-axis. The color key for both graphs is shown inset on the right-hand side. The letters A,B,C.. \textit{etc.} represent different neighborhoods for the relaxed Br interstitial. This figure demonstrates that E$_m$ is dependent not only on the initial and final structure local electronic environments but also on the transition structure local electronic environment.}
    \label{fig:interatomic_neighbors}
\end{figure*}

Our first task here was to identify possible interstitial sites for an extra bromine atom relaxed at different sites in the crystal.
We then looked at the neighborhood of these different relaxed bromine interstitials to map out possible sites.
Figure \ref{fig:interatomic_neighbors} lists the neighboring interatomic distances from the Br$_{i}$ atom for each \ce{CsPbBr3} relaxed structure without a paired vacancy. 
These correspond to different sites to which the interstitial Br relaxed in the supercell. 
The letters A,B,C.. \textit{etc.} represent different neighborhoods for the relaxed Br interstitial, further visualized in Figure S.2.1. 

\begin{figure}
    \centering
    \includegraphics[width = \textwidth/2]{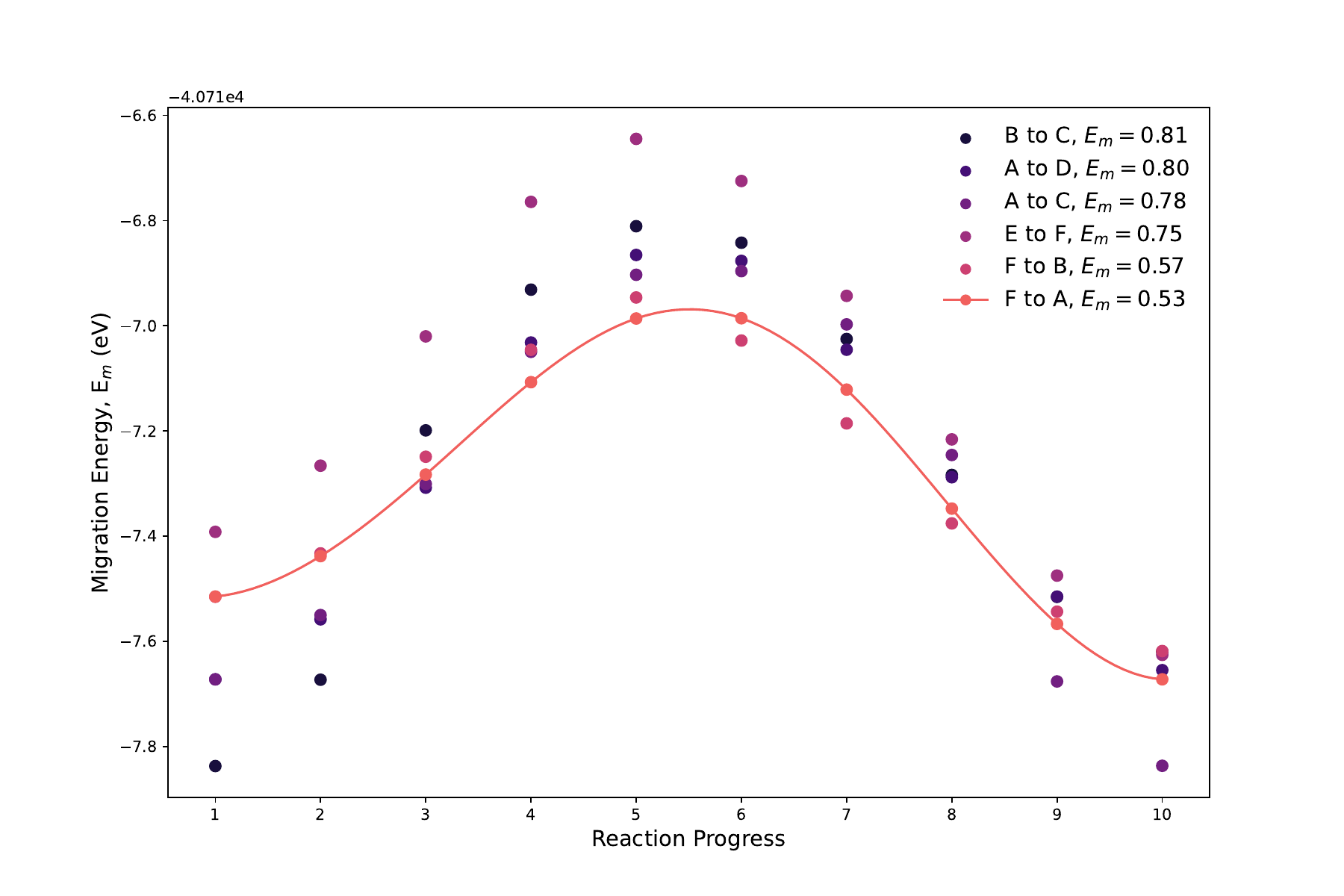}
    \caption{Neutral interstitial Br$_{i}$ E$_m$ (eV) in \ce{CsPbBr3} structures.}
    \label{fig:neutral_interstitial_NEB}
\end{figure}

A change in the coordinates of the pre-relaxation Br$_{i}$ of as little as 10\% in one direction resulted in different identities of its neighbors and corresponding distances between the reference Br$_i$ atom and the neighbors, as shown between systems C and D.\
We observed that, while Br$_i$ systems C and D exhibited different sets of neighbors after structural relaxation, the resulting migration energies from a reference system to systems C and D, respectively, were remarkably similar (E$_m$ = 0.78~eV for system A to system C, versus E$_m$ = 0.80~eV for system A to D) in Figure \ref{fig:neutral_interstitial_NEB}. Those differences are within the uncertainty of the predictions themselves. 
These small differences indicate that considering only the local electronic environment of the initial or final system does not fully explain the origin of the $E_m$ value. 
In contrast, systems C and D show similar transition state structures, \textit{i.e.,} at the saddle point, along the NEB pathway from A to C and A to D, respectively; Figure 3b.
Thus, the local Br$_i$ environment in both the initial structure \textit{and} the transition structure influences the value of E$_m$. 

Bringing all these works together, we calculate the more experimentally accessible activation energies for all the defects we studied.  Table \ref{tab:table3} reports the activation energy ($E_a$) associated with Br$_{i}$ returning to a vacant lattice site through direct recombination (creating a perfect crystal).

\subsection{Potential Mechanisms for Defect Migration}

We studied NEB calculations of the movement of interstitial sites occurring via two distinct mechanisms:  The first involved the assistance of vacancies in the lattice, while the second relied on a coordinated movement of the interstitials themselves.

\subsubsection{Vacancy-Assisted Mechanism}
To delve deeper into the intricacies of migration dynamics, we strategically introduced a vacancy into the lattice structure along with an interstitial. 
This approach allowed us to quantify the activation energies necessary for an interstitial atom to migrate towards, and subsequently recombine with, the vacancy (resulting in a healed perfect crystal).
Table \ref{tab:table3} reports the activation energy ($E_a$) associated with Br$_{i}$ returning to a vacant lattice site through direct recombination.

\begin{figure*}
    \centering
    \includegraphics[width = \textwidth]{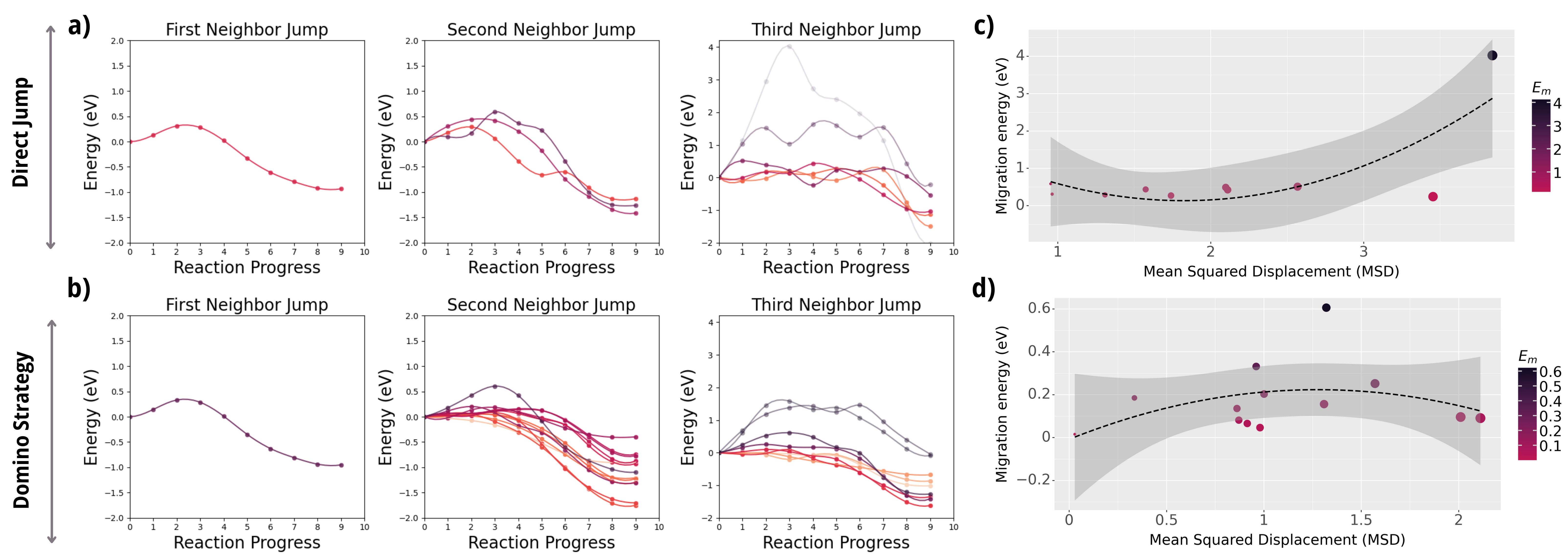}
    \caption{Migration energy of a \ce{Br_i-Pb_i} split interstitial following two distinctly different pathways and its effect on the first, second and third nearest neighbors of the moving Br atom (first three columns, respectively).  
    (a, top) The energy barrier for the migration of a \ce{Br_i-Pb_i} split interstitial via a direct jump to a neighboring vacancy site. Energy profiles for the first, second, and third nearest neighbors are grouped based on distance to the Br interstitial site. Each line stands for a different migration pathway which is color-coded by their activation rank order. 
    (b, bottom) The energy barrier associated with deploying a ``domino effect'' strategy for inter-lattice diffusion. 
    (c) Maps the correlation between activation energy and mean squared displacement (MSD) for the direct jump mechanism. Color-coded dots, based on activation energy, are rank-ordered with shaded areas indicating confidence intervals for the regression curve. This shows that E$_a$ increases with MSD, as expected. 
    (d) Displays a similar correlation to (c), except that this relates to the ``domino effect'' diffusion mechanism. Here, unlike in (c), the activation energy is pretty flat regardless of the MSD value.}

    \label{fig:activation_barrier_reduction}
\end{figure*}

In the context of vacancy-assisted migration, our investigations unveiled a ``domino effect'' strategy that can assist the translocation of an interstitial across the lattice, a phenomenon particularly pronounced in interactions involving second and third nearest neighbors. 
Rather than migrating directly into the vacancy site, the interstitial atom induces a successive movement of one or two adjacent atoms along the pathway, effectively achieving the final state through a series of intermediate steps.
This phenomenon is clearly demonstrated in Figure \ref{fig:activation_barrier_reduction}, which illustrates how the NEB-generated activation energy can be 'flattened' by this domino effect, providing an easier migration pathway for the atoms.
In Figure \ref{fig:activation_barrier_reduction}, we can clearly see that the barrier for long-distance jumps of the interstitial was significantly reduced by the domino effect. We are unaware of any other references to such a novel and low-energy pathway in the literature, marking this out as an important finding. 

We calculated the average mean square displacement (MSD) of each atom from its lattice site. This was computed for each pathway at the saddle point of diffusion moves that involve recombination of defects. 
Looking at the MSD for (a) and (c) in Figure \ref{fig:activation_barrier_reduction},  we can see that it follows a linear regime in the direct jump scenario: The activation energy increases with increasing MSD as might be expected. 
In contrast, in the domino effect strategy, (b) and (d) in Figure \ref{fig:activation_barrier_reduction},  the activation energy tapered off for sufficiently large MSD values of the interstitials. 
This shows that, for any migration more than a second or a third neighbor distance away, a domino movement is far more probable than a direct jump. 

A summary of the NEB findings can be found in Figure \ref{fig:Recombination_energy_summarized}, offering insights into the energetics of the vacancy-assisted migration process.

We examined, in some detail, one of the recombination pathways that involved a third neighbor jump. Here, we start from the direct jump and sequentially add one atom at a time to see how the number of intermediate atoms affects the final migration energy.
As depicted in Figure \ref{fig:activation_barrier_reduction}, the direct jump has two local energy peaks in the migration process, indicating two relatively high energy barriers to overcome. When an intermediate atom is introduced, only one peak remains in the migration pathway and the migration energy is reduced. We posit that the involvement of an intermediate atom assisted in helping the original interstitial atom bypass one of the barriers, although one barrier still remains. Introduction of yet another intermediate atom flattens the left of the two barriers and the resulting migration energy is effectively reduced to zero. This suggests that, even if the defect is relatively distant from the vacancy site, it can still migrate and recombine simultaneously. This suggests that some defects can self-heal without additional energy input if the ``domino effect'' strategy can be engaged.

\label{subsubsec:vacancyassisted}

Table \ref{tab:table3} reports the activation energy ($E_a$) associated with Br$_{i}$ returning to a vacant lattice site through direct recombination to a perfect crystal.

\begin{figure}
    \centering
    \includegraphics[width = \textwidth/2]{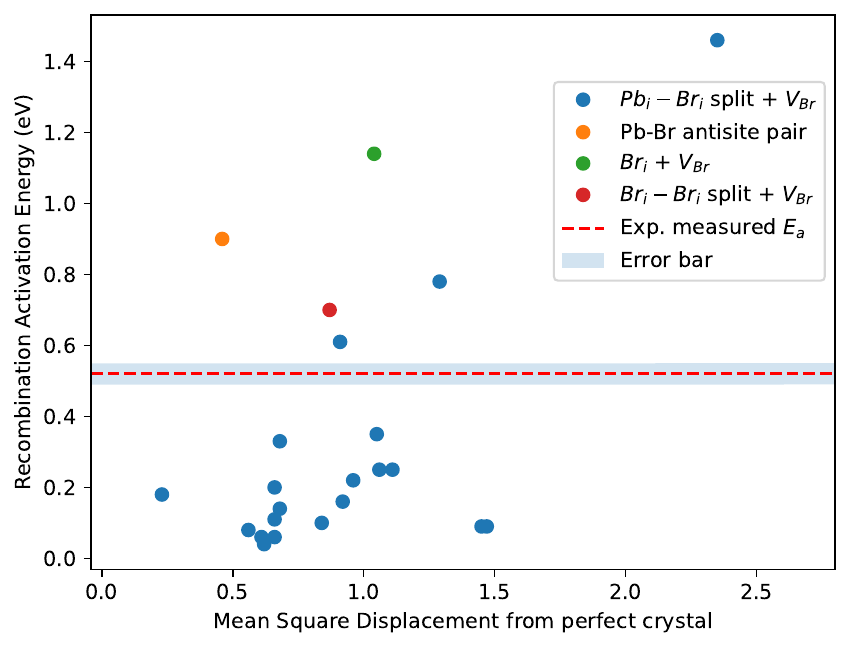}
    \caption{Summary of the activation energy values in this study for defect recombination events plotted as a function of the Mean Squared Displacement for a perfect crystal involved in the pathway and type of defect studied. Color key for the data shown is given on the top RHS of the figure. The plot shows that forming a variety of Pb-Br split interstitials in the presence of a Br vacancy (blue dots) offers many opportunities for low activation energy events (< 0.4~eV). The single experimental activation energy for CsPbBr$_3$ from Tirmzi \textit{et al.} is shown as a red dashed line with the error bar for measurement shown as shaded blue.}
    \label{fig:EvsMSD}
\end{figure}

\begin{figure}
    \includegraphics[width = \textwidth/2]{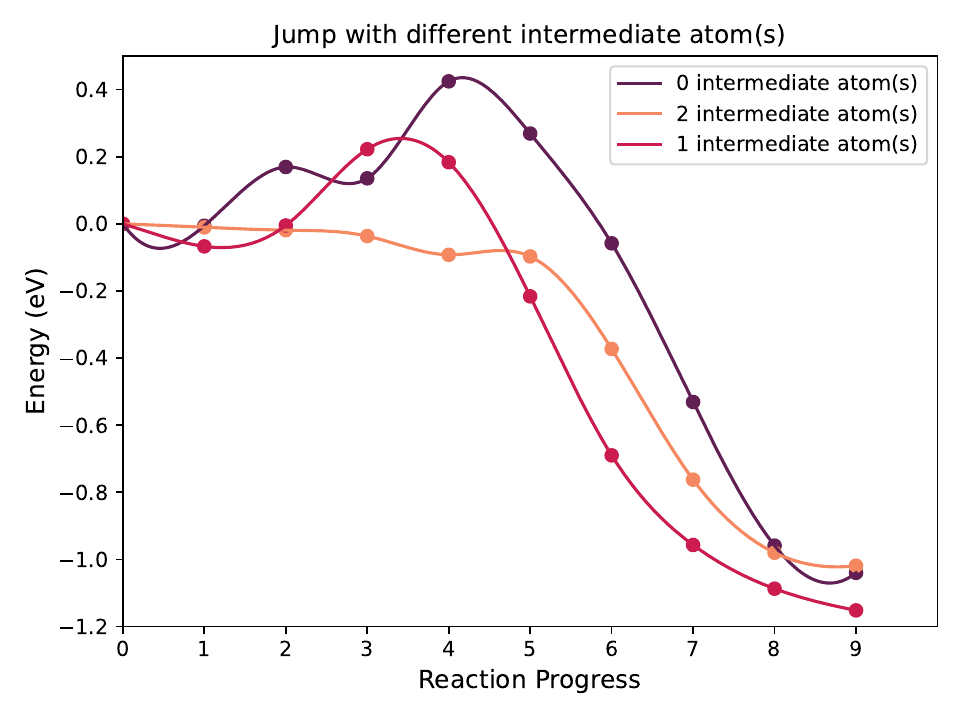}
    \caption{Migration energy differences for \ce{Br_i} migration with one (red line) or two (orange line) intermediate \ce{Br} atoms participating in the motion compared to zero atoms in the defect transition (purple line). The differences in migration energies are lower as one, and then two, intervening atoms are involved, representative of the ``domino effect'' strategy.}
    \label{fig:Em_40}
\end{figure}

\subsubsection{Interstitial Migration Mechanism}

In addition to the previously described \textit{vacancy-assisted} migration, our investigations have revealed that \ce{Br} interstitials are quite capable of migrating to alternate interstitial sites. 
This migration energy is influenced by the local environment of the interstitials in all three structures (initial, transition and final) as we saw in Section~\ref{defect_migration}. 
Notably, when the vacancy is not the closest neighbor to the interstitial, the ``domino effect'' strategy appears to facilitate this migration as well. 

\begin{figure*}[p]
    \centering
    \includegraphics[width=\textwidth]{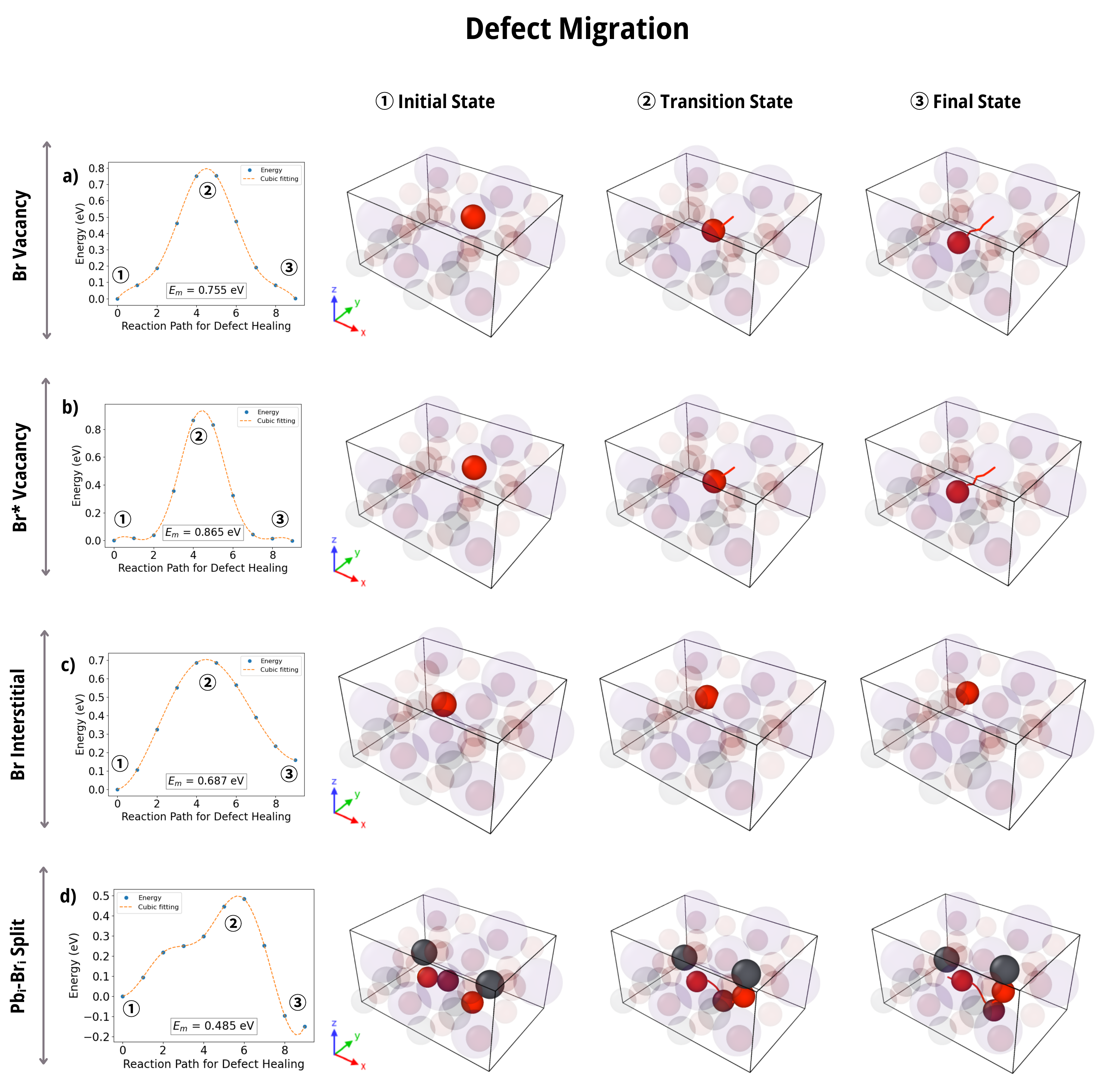}
    \caption{Summary of Defect Migration Energies for four defects: From top to bottom, a neutral Br vacancy, a positively charged Br vacancy, a Br interstitial (no vacancy) and a split interstitial involving lead and bromine atoms. Left column: The NEB-generated energy barrier.  
    The right three columns show a visualization of the initial state, the transition state, and the final state, as indicated. Atoms not engaged in the pathway are shown with reduced intensity. Those actively engaged in the pathway are shown in more intense colors. Key: Red = Br; Black = Pb. In all images, the red tail, shown as a solid line, indicates the path taken by the Br atom during the pathway}
    \label{fig:Migration_Energy_summarized}
\end{figure*}

\begin{figure*}
    \centering
    \includegraphics[width=\textwidth]{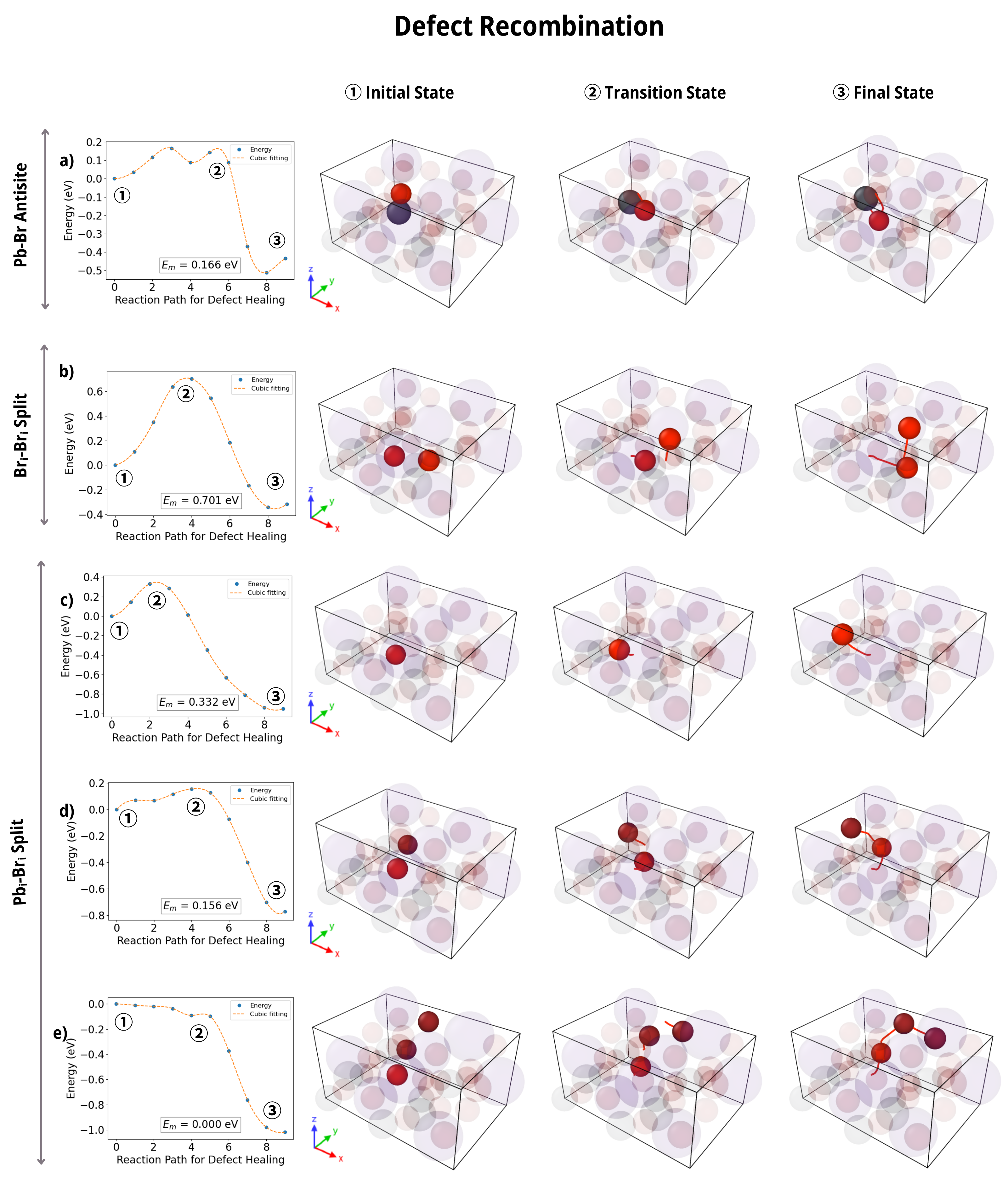}
    \caption{Summary of Recombination Energies for five defects: From top to bottom, a Pb-Br anti-site, a Br-Br split interstitial, and three Pb-Br split-interstitials that are aligned in the $x-$, $y-$ and $z-$ directions, respectively in the lattice. Left column: The NEB-generated energy barrier. The right three columns show a visualization of the initial state, the transition state, and the final state, as indicated. Atoms not engaged in the pathway are shown with reduced intensity. Those actively engaged in the pathway are shown in more intense coloring. Key: Red = Br; Black = Pb. In all images, the red tail, shown as a solid line, indicates the path taken by the Br atom during the pathway}
    \label{fig:Recombination_energy_summarized}
\end{figure*}

The NEB findings, in the form of migration energy barriers and a visualization of the pathway mechanism, as captured in the initial, transition, and final states of the system, are summarized in Figure \ref{fig:Migration_Energy_summarized}.
This Figure offers a compact overview and associated insight into the energetics and mechanisms of the defect migration process. 
These results clearly indicate that the most favorable migration pathway amongst the ones we studied features a collective and concerted movement of several atoms, consistent with the interstitial ``domino effect'' in vacancy-assisted migration.

Figure\ref{fig:Em_40} accurately captures this lowering of the activation energy through  the ``domino effect.'' 
We see that, for an interstitial to migrate, having one more intermediate atom involved reduced the activation barrier for migration.
Adding a second intermediate atom reduced this barrier almost to zero. 
This shows that for any defect travel larger than the nearest neighbor, the migration barrier can be reduced by a domino effect of atoms moving in a concerted exchange for metal halide perovskites.

\begin{table} 
\caption{\label{tab:table1}Defect Formation Energies in \ce{CsPbBr3}}
\begin{tabular*}{0.48\textwidth}{@{\extracolsep{\fill}}lcc}
\hline
Defect\footnote[4]{Br-rich Environment}&Formation Energy (eV)&Other Work (eV)\\
\hline
\ce{V_{Br}}&2.71&2.67\cite{Kang2017HighCsPbBr3}\\
\ce{Br_{i}}&0.75&0.70\cite{Kang2017HighCsPbBr3}\\
\ce{Pb_{Br}}&6.08&6.28\cite{Kang2017HighCsPbBr3}\\
\ce{Br_{Pb}}&1.30&1.40\cite{Kang2017HighCsPbBr3}\\
\ce{Br_i-Br_i} split&1.55& N/A \\
\ce{Br_i-Pb_i} split&1.19& N/A\\
\ce{Cs_{Pb}} & 0.67 & 0.81\cite{Kang2017HighCsPbBr3}\\
\ce{Pb_{Cs}} & 2.87 & 2.44 \cite{Kang2017HighCsPbBr3}\\
\hline
\end{tabular*}
\end{table}

\begin{table} 
\caption{\label{tab:table2}Defect Migration Energies in \ce{CsPbBr3}}
\begin{tabular*}{0.48\textwidth}{@{\extracolsep{\fill}}lc}
\hline
Defect Pathway&Migration Energy (eV)\\
\hline
\ce{V_{Br}} &0.89\\
\ce{V^._{Br}}&0.75\\
Br$_{i}$&0.53-0.81\\
\ce{Br_i-Br_i} split&0.62\\
\ce{Br_i-Pb_i} split&0.48\\
\hline
\end{tabular*}
\end{table}

\begin{table} 
\caption{\label{tab:table3}Activation Energies for Defect Recombination in \ce{CsPbBr3}}
\begin{tabular*}{0.48\textwidth}{@{\extracolsep{\fill}}lc}
\hline
Recombination Pathway&Activation Energy (eV)\\
\hline
Br$_{i}$ & 1.14\\
\ce{Pb_{Br}= \ce{Br_{Pb}}}& 0.17\\
\ce{Br_i-Br_i} split&0.70\\
\ce{Br_i-Pb_i} split&0.10$- $0.14\\
\hline
\end{tabular*}
\end{table}

A summary of the NEB-derived formation energies, migration energies and activation energies investigated in this work can be found in Tables \ref{tab:table1},\ref{tab:table2} \& \ref{tab:table3}, respectively.\footnote[4]{Br-rich environment}

\section{Methodology}

Nudged Elastic Band (NEB) calculations are a well-established methodology for finding the minimum energy pathway that follows the transition of one arrangement of atoms (the initial state) to a different arrangement (the final state). \cite{Mills1994QuantumSystems, BerneClassicalScientific, Henkelman2000ClimbingPaths, Henkelman2000ImprovedPoints}
These calculations give rise to an estimation of the maximum potential energy needed to effect this transition, the ``saddle point energy,'' which is assumed to be synonymous with the activation energy barrier for the transition. 

In the calculations here, we estimate both the defect's \textit{formation energy} and \textit{migration energy}, the energy required to move the defect from point A to point B in the supercell. 
Both of these energy barriers contribute to the overall activation energy for CsPbBr$_3$, as explained in the Introduction.\ 
We use a $3 \times 3 \times 2$ supercell of CsPbBr$_3$ containing a total of 360 atoms for the calculation of defect \textit{formation} energies. 
This is close to the practical limit of system size for a reasonably accessible computational resource effort. 
Perdew-Burke-Ernzerhof (PBE) calculations were performed using ultrasoft pseudopotentials\cite{Vanderbilt1990SoftFormalism} with a plane wave cut-off of 40 Ryd (400 Ryd on the charge density).
These pseudopotentials were chosen because they accurately represent defects in these kinds of systems.\cite{Meggiolaro2018First-PrinciplesIssues}
The well-known electronic structure code, Quantum ESPRESSO, \cite{Giannozzi2009QUANTUMMaterials, Giannozzi2017AdvancedESPRESSO, Giannozzi2021QuantumExascale} was used for all density functional theory (DFT) calculations reported below.  
The \textit{migration energies} and pathways for recombination of these defects were estimated using NEB. 
Due to the high computational cost of running Nudged Elastic Band simulations, we chose a smaller $2 \times 1 \times 1$ supercell of 40 atoms for the migration energy and the activation barrier for recombination. 
Details of all the parameters used to set up the Quantum ESPRESSO calculations, as well as sample scripts, are provided in the Supplementary Information.

Based on Kang \textit{et al.}'s work and that of others, we selected point defects that are thought to be particularly prevalent in lead halide perovskites. 
Some were chosen for their low defect formation energies.\cite{Kang2017HighCsPbBr3}
While much of our focus involved looking at simple point defects (V, I), we also explored the migration and recombination of additional ones not considered previously, including ``split interstitial'' defects in which two atoms share a lattice site, as well as more concerted pathways.

In the \textit{ab initio} calculations of defect energetics, we introduced a specific defect into a perfect lattice, fixed the cell parameters,  and then allowed the atoms to relax. The workflow for such calculations is shown in Figure~\ref{fig:flowchart}. 
The defect formation energy (DFE) was then computed based on the total energies of the relaxed structures, using the following equation: \cite{Wang2017AbInGaAs}
\begin{equation} \label{eq:2}
E^f(D^q) = E(D^q) - E_{\text{bulk}} - \sum_i \Delta n_i \mu_i + q(E_V + E_F) + E_{\text{corr}},
\end{equation}
where $E^f(D^q)$ represents the defect formation energy of a defect $D$ in charge state $q$ and $E_{\text{bulk}}$ the total energy of the pristine bulk supercell. 
This equation \ref{eq:2} incorporates the changes in the number of atoms ($\Delta n_i$) \& the chemical potentials ($\mu_i$). 
The Fermi level ($E_F$) of a semiconductor is treated as an independent variable that can assume any value within the band gap ($E_V$), the
valence band maximum (VBM) of the bulk material.
The correction term ($E_{\text{corr}}$) for finite size effects and periodic boundary conditions.

For the NEB calculations of the migration energy, we began by choosing initial and final configurations of a defective lattice and allowed them to relax. 
We created ten images between the initial and final states for each NEB-derived energy barrier to investigate a manually selected recombination or migration defect pathway.

As well as studies of simple vacancies and interstitials and anti-site defects, we also considered some more exotic, concerted defect pathways. A particularly interesting example involves a defect that ``moves'' in a concerted manner involving multiple atoms. It moves such that the interstitial atom you start with is not the same one that eventually moves across the lattice --- a ``domino effect.'' As we shall show, having other bromine interstitial atoms participate in the migration mechanism drastically reduces the associated activation barrier. A more comprehensive explanation of the ``\textit{domino effect}'' in \ce{CsPbBr3} can be found in the Supplementary Information. 
We identified the intermediate atoms along the ``domino effect'' pathway in the NEB calculation and compared the migration energy with that of a direct jump, as well as a second neighbor jump (involving two Br atoms) and a third neighbor jump (involving three Br atoms). These comparisons allowed us to examine the contribution of each additional atom to the change in migration energy.

We now consider the creation of a set of defective systems studied in this paper.

\renewcommand{\thesubsection}{\arabic{section}.\arabic{subsection}}
\subsection{Br vacancies}

Halide point defects, especially vacancies, are expected to be one of the most, if not the most, readily moveable defect in MHPs.\cite{Azpiroz2015DefectOperation,Kaiser2022DefectPerovskites}
As there is a single unique site for bromine atoms, a vacancy can be formed by removing a bromine atom from the simulation cell and letting the system relax. 
We also looked at the effect of migration of a vacancy in both a charged and a neutral supercell. 
The migration of a vacancy to its nearest neighbor site and the resulting activation energy associated with the creation and migration of these vacancies are given in Table \ref{tab:table1}.

\subsection{Br interstitials}
To create an interstitial Br in the system (Br$_{i}$), we placed a Br atom at a location offset from a bulk crystal reference site by a user-specified amount, typically 10-25\% in the $x$-, $y$-, $z$- directions,  to test the tolerance of the resulting relaxed configuration to its starting point. Then we performed a \textit{relax} calculation on the system.  This is a different approach to other ways to predict the Br${_i}$ locations, \textit{e.g.,}  from a Voronoi analysis or other methods. \cite{Goyal2017ACalculations} 
We characterized six unique interstitial locations, determined by their differing nearest neighbor distances and atom types, Figure \ref{fig:interatomic_neighbors}. 
Their reference coordinates can be found in the Supplementary Information, Table S2.1.
To determine E$_m$, different Br$_{i}$ systems were selected to be the first and final NEB images, with the resultant ``forward'' direction selected as the lower E$_m$ value. 
For recombination, the activation energy ($E_a$) is the energy associated with Br$_{i}$ returning to a vacant lattice site through direct recombination to create a perfect crystal.

\subsection{Split interstitials}
A split interstitial consists of two self-interstitials sharing a single substitutional lattice site.  These are some of the most common interstitials found in semiconductor materials, \cite{Masri1983NatureSilicon} especially amongst the halide defects\cite{Park2019AccumulationPerovskites}. 
We explored the defect migration energies of split interstitials involving two bromine atoms sharing one site (Br-Br) and a bromine and lead atom sharing a site (Br-Pb). 
Both of these are located at the bromine lattice sites. 
Table \ref{tab:table2} summarizes the formation energy for creating these split interstitials.

\subsection{Anti-sites}
An anti-site refers to the presence of a ``foreign'' (unexpected) atom on a lattice site. 
Here, we explore two types of anti-sites: The first involves lead on a bromine site, and bromine on a lead site (\ce{Pb_{Br} and Br_{Pb}}), followed by simulation of the recombination of a pair of anti-sites \ce{Br_{Pb}}= \ce{Pb_{Br}} into the null (perfect) crystal system. The second involves Pb on a Cs site, and vice versa.  

The atom sites were carefully selected to ensure that they are located at the center of the supercell. 
To simulate the recombination of a pair of anti-sites into the bulk crystal, the two displaced atoms had to be introduced within a first nearest neighbour distance. Twenty such candidate sites were identified. 
Amongst these, the lowest energy system was selected to further investigate the recombination of a pair of anti-sites into the perfect crystal.
Defect formation energies were calculated for single anti-site defect and tabulated in Table \ref{tab:table1}.

\section{Conclusions}

We have studied a broad range of defects in the \ce{CsPbBr_3} metal halide perovskite, including the defect formation energies (E$_f$) and migration energies (E$_m$) studied using accurate \textit{ab initio} Nudged Elastic Band calculations. The latter will be a helpful addition to the literature. The formation energies reproduce those of Kang and Wang \cite{Kang2017HighCsPbBr3}. 
Unlike other studies, the combination of both E$_f$ and E$_m$ allows us to calculate the activation energy of the various defects, which could be compared to experiment.
Above, we highlighted Tirmzi \emph{et al.}'s experiment measuring the slow, activated recovery of light-induced conductivity in \ce{CsPbBr3}. 
We find that the calculated $E_m$ for $\mathrm{V}_{\mathrm{Br}}^\bullet$ agrees well with the 0.53~eV activation energy for relaxation of light-induced conductivity observed by Tirmzi and co-workers.
Beyond the more well-studied intrinsic defects, namely, a Br vacancy and a Br interstitial defect, we have included additional defects like the family of split interstitials (Br-Br and Pb-Br) as well as anti-site defects. Finally, we have identified a formerly unexplored diffusion pathway which we have labeled the ``domino effect'' strategy involving multiple proximal Br interstitials. 
The merits of this latter pathway are described below.

The low activation energies that we observed for some of the recombination mechanisms can help explain the self-regulation process seen in perovskites. \cite{Yao2021EnhancedDefects}   
Defects like vacancies, interstitials and anti-sites help the perovskite ``heal'' by adjusting the concentration of charge carriers. 
The pathways that the 10-window NEB calculations allowed us to explore give us accurate energy barriers as well as insight into the different mechanisms of self-regulation that metal halide perovskites can undergo. 
The domino effect of recombination aided by other nearby bromine atoms lowered the activation threshold, a phenomenon we saw across most of the pathways we studied.

Looking at a holistic picture of all the recombination pathways we studied and their associated energy barriers, the lowest value of the activation energy that we observed (see Figure \ref{fig:EvsMSD}) was 0.10~eV and a number of Pb-Br split interstitials exhibited energy barriers below 0.4~eV.  
Br-Br split interstitials in the presence of a vacancy can also heal with a barrier under 0.7~eV. 
An experimentally measured activation energy will be an ensemble average of a number of pathways, and must surely include many of the ones we have uncovered here. 

Investigating the pathway that results in the lowest barrier (0.1~eV), this mechanism is governed by the process \ce{V_{Br} + Br_i-Pb_i split -> \ce{Br_{Br} + Pb_{Pb}}}, with the interstitial being located at the third nearest neighbor to the vacancy and the reaction being aided by two other proximal bromine atoms in the lattice. 
Because of its low formation energy and the ease of movement for split interstitial defects in the system, the activation energy is lowered by quite a large margin.
Adding to this lower energy, the participation of other bromine atoms to create a ``domino effect'' drastically decreases the energy barrier by reducing the total displacement that each bromine has to travel to return to the defect-free bulk state.
As the defect moves, it affects other intermediary atoms, transferring energy and reducing the overall activation energy needed. 
Essentially, the intermediary atom acts as a stepping stone, making defect movement more efficient.
This ``domino effect'' strategy, involving a collective movement of defects in inorganic metal halide perovskites, like CsPbBr$_3$, enables a drastic reduction of the activation energy needed to traverse the lattice.  

The calculations of defect pathways as well as E$_f$, E$_m$, and E$_a$ values provided in this paper can have an impact on the design of better materials and, ultimately, devices. Here are two pertinent suggestions:  First, the links we provide between structure and their corresponding formation energies can assist the the interpretation of X-ray diffraction patterns. While challenging, it is possible to use this paper's data to uncover the type and concentration of defects that correspond to a particular choice of processing conditions. The conjunction of this paper's data and the XRD patterns could be used to guide the design of processing protocols to improve defect management which is a key design feature for the future for perovskites\cite{young2018controldefect}.

Second, this paper has suggested that there are low-energy pathways that can self-heal defects. How can we exploit this observation? We hypothesize that proximal (NN to 3rd NN) defects generated by illumination will readily heal. It is uncertain what will transpire in more defective systems, putting a premium on starting with more perfect crystals. This is a testable hypothesis, experimentally and computationally, which we will pursue in future work.

\section*{Author Contributions}
\textbf{Kumar Miskin} : conceptualization, methodology, data curation, visualization and writing. \textbf{Yi Cao, Madaline Marland, Farhan Shaikh}: data curation, visualization and writing. \textbf{Jay Rwaka} : data curation.  \textbf{David Moore} : funding acquisition, writing - reviewing and editing. \textbf{John Marohn} : conceptualization, funding acquisition and writing. \textbf{Paulette Clancy} : conceptualization, funding acquisition, project administration, resources, supervision and writing.

\section*{Data Availability}
All scripts as detailed in SI can be found on the github link : \url{https://github.com/pclancy-lab/perovskite_defect_study/}

\section*{Conflicts of interest}
There are no conflicts to declare.

\section*{Acknowledgements}
The authors acknowledge support from the U.S. Department of Energy (DOE) Basic Energy Sciences award DE-SC0022305. 
Miskin thanks Johns Hopkins University for his support in AY 22-23.
Computational resource support was provided by the petascale Hopkins facility, Advanced Research Computing at Hopkins (ARCH) (rockfish.jhu.edu), supported by National Science Foundation award OAC 1920103. Partial funding for ARCH's infrastructure was originally provided by the State of Maryland.



\balance



\bibliography{references} 
\bibliographystyle{acm} 

\end{document}